\documentstyle[lprocl,colordvi,epsfig,11pt]{article}
\textheight
8.5in  \parskip 7pt \openup1\jot 
\def\st{\scriptstyle}

\def\be{\begin{equation}}
\def\ee{\end{equation}}
\def\bea{\begin{eqnarray}}
\def\eea{\end{eqnarray}}

\newskip\humongous \humongous=0pt plus 1000pt minus 1000pt
\def\caja{\mathsurround=0pt}
\def\eqalign#1{\,\vcenter{\openup1\jot \caja
        \ialign{\strut \hfil$\displaystyle{##}$&$
        \displaystyle{{}##}$\hfil\crcr#1\crcr}}\,}
\newif\ifdtup

\def\eqright #1\cr{\noalign{\hfill$\displaystyle{{}#1}$}}
\def\eqleft #1\cr{\noalign{\noindent$\displaystyle{{}#1}$\hfill}}

\def\oldreffmt#1{\rlap{[#1]} \hbox to 2\parindent{}}

\def\figfmt#1{\rlap{Figure {#1}} \hbox to 1in{}}

\def\auto{\eqno(\refstepcounter{equation}\theequation)}
\def\begineq #1\endeq{$$ \refstepcounter{equation}\eqalign{#1}\eqno
	(\theequation) $$}
\def\contlimit{\,{\hbox{$\longrightarrow$}\kern-1.8em\lower1ex
\hbox{${\scriptstyle (a\rightarrow0)}$}}\,}
\def\centeron#1#2{{\setbox0=\hbox{#1}\setbox1=\hbox{#2}\ifdim
\wd1>\wd0\kern.5\wd1\kern-.5\wd0\fi
\copy0\kern-.5\wd0\kern-.5\wd1\copy1\ifdim\wd0>\wd1
\kern.5\wd0\kern-.5\wd1\fi}}
\def\centerover#1#2{\centeron{#1}{\setbox0=\hbox{#1}\setbox
1=\hbox{#2}\raise\ht0\hbox{\raise\dp1\hbox{\copy1}}}}
\def\centerunder#1#2{\centeron{#1}{\setbox0=\hbox{#1}\setbox
1=\hbox{#2}\lower\dp0\hbox{\lower\ht1\hbox{\copy1}}}}
\def\lsim{\;\centeron{\raise.35ex\hbox{$<$}}{\lower.65ex\hbox
{$\sim$}}\;}
\def\gsim{\;\centeron{\raise.35ex\hbox{$>$}}{\lower.65ex\hbox
{$\sim$}}\;}
\def\st#1{\centeron{$#1$}{$/$}}

\def\super#1{\ifmmode \hbox{\textsuper{#1}}\else\textsuper{#1}\fi}
\def\textsuper#1{\newcount\holdspacefactor\holdspacefactor=\spacefactor
$^{#1}$\spacefactor=\holdspacefactor}

\def\getcite#1,{\advance\citenumber by1
\ifnum\citenumber=1
\ref{#1}\let\next=\getcite\else\ifx#1@\let\next=\relax
\else ,\ref{#1}\let\next=\getcite\fi\fi\next}

\def\upon #1/#2 {{\textstyle{#1\over #2}}}
\relax

\def\til#1{\centeron{\hbox{$#1$}}{\lower 2ex\hbox{$\char'176$}}}
\def\tild#1{\centeron{\hbox{$\,#1$}}{\lower 2.5ex\hbox{$\char'176$}}}
\def\sumtil{\centeron{\hbox{$\displaystyle\sum$}}{\lower
-1.5ex\hbox{$\widetilde{\phantom{xx}}$}}}

\def\pom{{\rm P\kern -0.53em\llap I\,}}
\def\spom{{\rm P\kern -0.36em\llap \small I\,}}
\def\sspom{{\rm P\kern -0.33em\llap \footnotesize I\,}}

\newfont{\unc}{eurb10} 
\def\q{\unc q} 

\begin{document} 

\begin{titlepage} 

\rightline{\vbox{\halign{&#\hfil\cr
&ANL-HEP-CP-99-112 \cr
&\today\cr}}} 
\vspace{1.25in} 

\begin{center}
{\bf THE TRIANGLE ANOMALY IN THE TRIPLE-REGGE LIMIT}\footnote{Work 
supported by the U.S.
Department of Energy, Division of High Energy Physics, \newline Contracts
W-31-109-ENG-38 and DEFG05-86-ER-40272} 
\medskip

Alan. R. White\footnote{arw@hep.anl.gov }

\end{center}
\vskip 0.6cm

\centerline{High Energy Physics Division}
\centerline{Argonne National Laboratory}
\centerline{9700 South Cass, Il 60439, USA.}
\vspace{0.5cm}

\begin{abstract} 
The U(1) triangle anomaly is present, as an infra-red
divergence, in the six-reggeon triple-regge interaction vertex obtained from
a maximally non-planar Feynman diagram in the full triple-regge limit of
three-to-three quark scattering. 
\end{abstract} 

\vspace{2in}
\begin{center}

Presented at the XXIX International Symposium on Multiparticle Dynamics

``QCD \& Multiparticle Production'', 

Brown University, Providence, August 9-13, 1999.

\end{center}

\end{titlepage}

\section{Background Motivation} 
Our eventual goal~\cite{arw98} is to find a Reggeon Field Theory (RFT) 
description of the pomeron in QCD. For this the pomeron should be 
a Regge pole plus multi-pomeron exchanges and interactions - in
agreement with experiment but not with perturbative QCD.
A major motivation is that the RFT Critical pomeron~\cite{cri} is the only
known asymptotic solution of both $s$- and $t$- channel unitarity producing
a rising total cross-section. In addition we also anticipate that 
the fundamental properties of 
confinement and chiral-symmetry breaking will be related to the 
regge behavior of the pomeron.

We have previously established~\cite{arw1} the existence of 
a ``super-critical phase'' of RFT which contains a regge pole pomeron
but also contains both a reggeized massive vector (gluon?) 
partner for the pomeron and a ``pomeron condensate''. 
These are properties we might expect to find in a color superconducting
phase of QCD in which SU(3) color is broken to SU(2). The
symmetry-breaking vector mass would then be 
identified with the RFT order parameter and 
the critical pomeron would appear at the critical point where color 
superconductivity disappears and full SU(3) symmetry is restored.
Also, in this identification, the condensate should carry the 
quantum numbers of the winding-number current suggesting, perhaps, that it 
is associated with spectral flow of the Dirac sea. This is why we 
look for a reggeon condensate arising from an ``anomaly'' infra-red 
divergence which we will eventually study in the dynamical context of the 
superconducting phase.

We must first demonstrate that it is possible for the anomaly to actually 
appear in a 
regge-limit effective theory. It is,
of course, absent in the usual perturbation expansion of a vector theory.
This talk will concentrate on this issue. 
The consequences (RFT for the pomeron, confinement, etc.) will be discussed 
only very briefly. An extended version of the 
analysis we outline, together with other closely related analyses, 
can be found elsewhere~\cite{arw99}.

\section{A New Manifestation of the U(1) Anomaly.}
We study 3-3 quark scattering in a ``triple-regge'' limit~\cite{gw} 
involving three large 
light-cone momenta and consider diagrams containing a single quark loop.
\newline\parbox{2.7in}{
\begin{center}
\leavevmode
\epsfxsize=1.4in
\epsffile{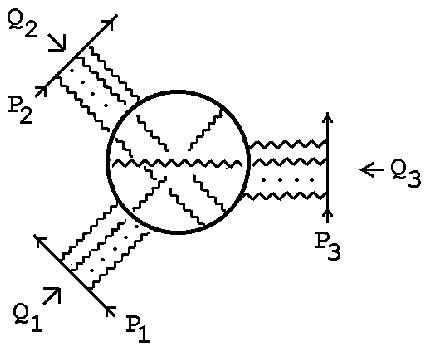}

Fig.~1 3-3 Quark Scattering

$~$
\end{center}
}
\parbox{3in}{ 
$$
\eqalign{ P_1~\to~ P_1^+~=& ~(p_1,p_1,0,0)~,~~p_1~ \to \infty \cr
P_2~\to~ P_2^+~=& ~(p_2,0,p_2,0)~,~~p_2~ \to \infty \cr
P_3~\to~ P_3^+~=& ~(p_3,0,0,p_3) ~,~~p_3~ \to \infty \cr
Q_1~\to&~~ (\hat{q}_1,\hat{q}_1,q_{12},q_{13})\cr
Q_2~\to&~ ~(\hat{q}_2,q_{21},\hat{q}_2,,q_{23})\cr
Q_3~\to&~~(\hat{q}_3,q_{31},q_{32},\hat{q}_3)}
$$}
\newline Using light-cone co-ordinates the leading behaviour (up to 
logarithms) is obtained by putting quark lines on-shell via $k_{i^{\pm}}$
integrations. As illustrated in Fig.~2 the resulting  
\newline \parbox{2.4in}{
``reggeon diagrams''~\cite{arw98} 
are $k_{\perp}$-integrals in which the ``reggeon interactions'' are quark
triangle diagrams with local (and non-local)
``effective vertices'' containing $\gamma$-matrix products which 
(in some cases) give the $\gamma_5 \gamma_{\mu}$ 
couplings that could produce the triangle anomaly. In particular 
the ``maximally non-planar'' diagram of Fig.~3 gives a 
six-reggeon interaction as shown.}
\parbox{3.4in}{
\begin{center}
\leavevmode
\epsfxsize=3in
\epsffile{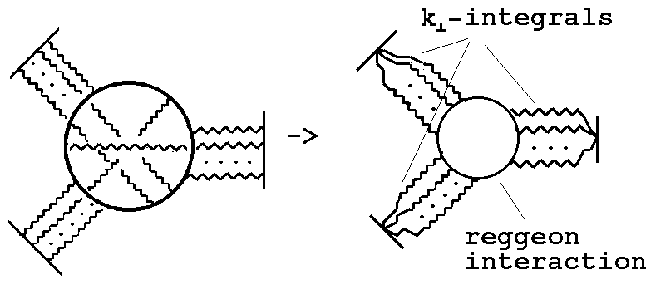}
\newline Fig.~2 Reggeon Diagram Generation.
\end{center}
}

\noindent \parbox{2.8in}{
\begin{center}
\leavevmode
\epsfxsize=1.8in
\epsffile{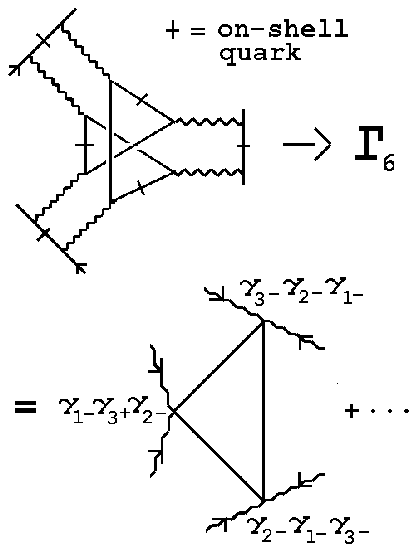}

Fig.~3 Six-Reggeon Interaction
\end{center}}
\parbox{3.1in}{ $~~~$
At lowest-order (two gluons in each $Q_i$ channel) there are 
$\sim 100$ diagrams that potentially give contributions.
To systematically evaluate all contributions 
and to discuss cancelations, 
it is necessary to develop a multi-regge 
asymptotic dispersion relation formalism~\cite{sw}
in which multiple discontinuities 
are initially calculated rather than amplitudes. Using this formalism it can 
be shown~\cite{arw99} 
that the anomaly occurs only in the maximally non-planar diagrams.
In this talk we will not discuss the dispersion relation formalism at all
but rather concentrate on showing how the anomaly occurs in such diagrams. 

To avoid ultra-violet subtleties 
we look for the anomaly in the infra-red region. 
}

\section{The Anomaly as an Infra-red Divergence.}
For a massless quark axial-vector current 
the anomaly divergence equation for the three current vertex 
$\Gamma_{\mu\nu\lambda}({\hbox{\q}}_{1}, {\hbox{\q}}_{2}, {\hbox{\q}}_{1} +
{\hbox{\q}}_{2})$ implies~\cite{cg}
$$
\Gamma_{\mu\nu\lambda} ~~
\sim  ~ \epsilon_{\mu\nu\alpha\beta} 
~{ {\hbox{\q}}_{1}^{\lambda}{\hbox{\q}}_{1}^{\alpha}{\hbox{\q}}_{2}^{\beta} 
\over {\hbox{\q}}_1^2 }
~+~ \cdots ~~,~~~~~~~ 
{\hbox{\q}}_1^2 \sim {\hbox{\q}}_2^2 \sim ({\hbox{\q}}_1 +
{\hbox{\q}}_2)^2  \sim {\hbox{\q}}^2 ~~\to 0
\auto
$$
For the U(1) current we are, of course, ignoring non-perturbative 
contributions. We will be looking for a ``perturbative'' effect! 
If ${\hbox{\q}}_{1+} \st{\to} 0~$, and ${\hbox{\q}}_2$ is spacelike with 
$ {\hbox{\q}}_2 \perp {\hbox{\q}}_{1+}$ 
then 
$$
\epsilon_{\mu\nu\alpha\beta} ~{\hbox{\q}}_{1}^{\lambda}{\hbox{\q}}_{1}^{\alpha}
{\hbox{\q}}_{2}^{\beta} 
/ {\hbox{\q}}^2 ~ \sim ~ {{\hbox{\q}}_{1+}}^2 /{\hbox{\q}} ~\sim ~ 1 
/{\hbox{\q}}  
\auto
$$ 
For our purposes, this linear divergence 
characterises the anomaly since we will not have a Lorentz-covariant amplitude
that separates into kinematic and invariant factors.
The divergence is produced by the triangle 
diagram Landau singularity and the light-like momentum ${\hbox{\q}}_{1+}$ 
is essential.
It can only be canceled by other quark triangle contributions.

\section{ A Contribution from a Maximally Non-planar Diagram.}

\noindent \parbox{2.3in}{
We label loop momenta as in Fig.~4. 
For the $k_1$ and $k_2$ integrations 
we use (new) light-cone co-ordinates~\cite{arw99} 
$$
k_i=k_{i2^-}\underline{n}_{1+}+ 
k_{i1^-}\underline{n}_{2^+}+\underline{k}_{i12+}
$$
$i=1,2$, together with conventional light-cone co-ordinates 
}
\parbox{4in}{ 
\begin{center}
\leavevmode
\epsfxsize=3.3in
\epsffile{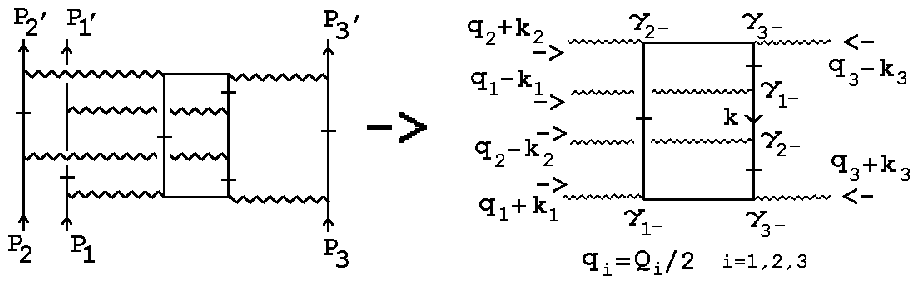}

Fig.~4 Loop Momenta
\end{center}
}
\newline ($k_{3^-},k_{3^-},
\underline{\tilde{k}}_{3\perp}$) for the $k_3$ integration. The 
$k_{11^-},k_{22^-}$, $k_{3^-}$ integrations are straightforward.
There are six 
options for using the remaining longitudinal $k_i$ integrations to 
put hatched lines on-shell. 
We look for the anomaly to be generated by a combination of 
local $\gamma$-matrix couplings, with an odd number of $\gamma_5$'s.
Only one option gives couplings, all three of which are local. A local
coupling is obtained from that part of a
quark numerator with the same 
momentum factor that scales the integrated longitudinal momentum, e.g.
$$
\eqalign{ &\int d k_{12^-} \delta\biggl( (k_1 + k - q_1)^2 - m^2 \biggr)
\gamma_{3^-} \biggl( (k_1 +k - q_1) \cdot 
\gamma + m \biggr)  \gamma_{1^-} \cr
&~~=~ \int d k_{12^-} \delta\biggl( k_{1^-}~ k_{12^-}
 + \cdots \biggr) \gamma_{3^-} \biggl( k_{1^-}  \cdot 
\gamma_{2^-} + \cdots \biggr) 
\gamma_{1^-} ~~ =~  \gamma_{3^-} \gamma_{2^-} \gamma_{1^-} ~ + ~\cdots \cr
}
\auto
$$

That part of the asymptotic amplitude with local couplings is
$$
\eqalign{& ~~~~~~~~~~~~~~ g^{12} ~~{p_{1}~p_{2}~p_{3} \over m^3} 
~~~~~\times \cr
&\int { d^2 \underline{k}_{112+} \over
(q_1 + \underline{k}_{112+})^2(q_1 - \underline{k}_{112+})^2}
 \int {d^2  \underline{k}_{212} \over
(q_2 + \underline{k}_{212+})^2(q_2 - \underline{k}_{212+})^2}
~\int  {d^2  \underline{k}_{33\perp} \over
(q_3 + \underline{k}_{33\perp})^2(q_3 - \underline{k}_{33\perp})^2} 
\cr  
&\int d^4 k~{ Tr \{ \hat{\gamma}_{12} (\st{k}+ \st{k}_1 
+ \st{q}_2 + \st{k}_3 +m) 
\hat{\gamma}_{31}( \st{k} +m) 
\hat{\gamma}_{23} (\st{k}- \st{k}_2 + \st{q}_1 
+ \st{k}_3 +m) \} \over 
([k + k_1 + q_2 + k_3]^2 - m^2) 
(k^2 - m^2)([k - k_2 + q_1 +k_3]^2 - m^2)} }
\auto
$$
where $k_{11^-}= k_{22^-} = k_{33^-}=0$, $~k_{12^-}, k_{21^-}$ and $k_{33^+}$ 
are determined by mass-shell $\delta$-functions, and 
$$
\eqalign{\hat{\gamma}_{31} ~&=~\gamma_{3^-}\gamma_{2^-}\gamma_{1^-}
~=~\gamma_{-,+,-}~-~ i~\gamma_5~
\gamma_{-,-,-} \cr
\hat{\gamma}_{23} ~&=~\gamma_{2^-}\gamma_{1^-}\gamma_{3^-}
~=~\gamma_{+,-,-}~-~ i~\gamma_5~
\gamma_{-,-,-}  \cr
\hat{\gamma}_{12} ~&=~\gamma_{1^-}\gamma_{3^+}\gamma_{2^-}
~=~\gamma_{-,-,-}~+~ i~\gamma_5~
\gamma_{-,-,+}  }
\auto
$$
with $ \gamma_{\pm,\pm,\pm} = \gamma\cdot n_{\pm,\pm,\pm}, ~~
n_{\pm,\pm,\pm} ~=~ (1,\pm1, \pm1, \pm1)$.

Removing the transverse momentum integrals and gluon propagators 
as the lowest-order 
contributions of two-reggeon states in each $t_i $-channel,
the three $\gamma_5$ couplings give the $m=0$ reggeon interaction 
$$
\eqalign{ &\Gamma_6(q_1,q_2,q_3,
\tilde{\underline{k}}_1,\tilde{\underline{k}}_2, 
\underline{k}_{3\perp},0) ~=\cr
&~~~~~ \int d^4 k  {  Tr \{ 
\gamma_5 \gamma_{-,-,+} (\st{k}+ \st{k}_1 + \st{q}_2 +\st{k}_3) 
\gamma_5 \gamma_{-,-,-} ~\st{k}~ 
\gamma_5 \gamma_{-,-,-}(\st{k}- \st{k}_2 + \st{q}_1 + \st{k}_3 ) 
\over  (k + k_1 + q_2 + k_3 )^2  
~k^2 ~
 (k - k_2 + q_1 + k_3)^2 } }
\auto
$$
\parbox{3.3in}{This interaction corresponds to the triangle diagram of Fig.~5. 
As discussed above, to obtain the maximal anomaly infra-red divergence
we must have a component of the axial-vector triangle diagram
tensor  $\Gamma^{\mu\nu\lambda}$  with
$\mu= \nu $ having a lightlike projection and $\lambda $ 
an orthogonal index having a spacelike projection.
This requirement is met if 
the light-like projection is made on either
$\underline{n}_{1^+}$ or $\underline{n}_{2^+}$.
In each case,
$\gamma^{-,-,+}$ has the necessary orthogonal spacelike component.
The anomaly divergence appears if we can take 
the limit
}
\parbox{2.5in}{\begin{center}
\leavevmode
\epsfxsize=1.6in
\epsffile{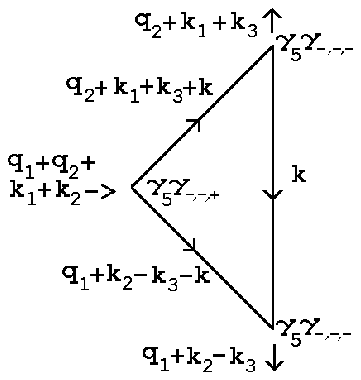}

Fig.~5 A Triangle Diagram. 
\end{center}
}
$$
(k_1 + q_2 +k_3)^2 \sim (q_1 + q_2 + 
k_1 +k_2)^2 \sim (k_2 + q_1 - k_3)^2 \sim 
\hbox{\q}^2 \to 0
\auto
$$
while keeping a finite light-like momentum, parallel to either
$\underline{n}_{1^-}$ or $\underline{n}_{2^+}$, flowing through the diagram. 
The mass-shell constraints must also be satisfied. 

We first consider $\hbox{\q}=0$ with
the loop momentum $k \sim \hbox{\q} = 0 $. It is then straightforward to add
momenta that are $O(\hbox{\q})$. If we impose 
$$
q_1 + q_2 + k_1 + k_2 ~= ~0~, ~~~~~q_1 + k_2 - k_3 ~= ~(2l,-2l,0,0)~
~\sim~ \underline{n}_{1^-}
\auto
$$
and take $q_1-k_1$ and $q_2-k_2$ lightlike with 
$$
q_{12^-} =-k_{12^-} = l~, ~~ q_{21^-} =-k_{21^-} =- l~,~~~~
\underline{q}_{112+} = \underline{k}_{112+}
=- \underline{q}_{212+} =- \underline{k}_{212+} 
\auto
$$
then $q_3 = -q_1 - q_2 = (0,l,-l,0)$ has the necessary spacelike form
and if $k_3=(0,3l,l,0)$ with 
$$
8lq_{112-}~=~q_1^2~=~\underline{q}_{112+}^2~=~q_{112-}^2 + q_{13}^2 
\auto
$$
then all requirements are satisfied and 
$$
(q_1+k_1)^2~=~4q_1^2~=~(q_2+k_2)^2~=~4q_2^2~=~32lq_{112-}
~=~{32 \over \sqrt{2}}(q_3^2)^{1/2}
(q_1^2 -q_{13}^2)^{1/2}
\auto
$$ 

Adding momenta $O(\hbox{\q})$ parallel to $\underline{n}_{12-}$, the limit 
$\hbox{\q}\to 0$, with $l$ fixed, gives 
$$
(q_1 -k_1)^2~\sim~(q_2 -k_2)^2 ~\sim~ \hbox{\q}^2 ~ \to 0~
\auto
$$
while 
$$
(q_1 +k_1)^2~ \to ~(q_2+ k_2)^2 ~\st{\to} ~0~,~~~ (q_3-k_3)^2 ~\to
~ 2(q_3+k_3)^2 ~\st{\to} ~0 
\auto
$$
and the six-reggeon vertex $\Gamma_6$ has the divergence
$$
\Gamma_6 ~\sim  ~
\epsilon_{\underline{n}_{1^+},\underline{n}_3,\underline{n}_{2^+},
\underline{n}_{12+}} 
~~{l^2 \over \hbox{\q}} 
~~\sim ~~ { Q_3^2 \over \hbox{\q} }
\auto
$$
If we instead impose 
$$
q_2 + k_1 + k_3 ~= ~(2l,0,-2l,0) ~~\sim~ \underline{n}_{2^-} 
\auto
$$
the role of $1$ and $2$ is interchanged together with 
$k_3 \leftrightarrow - k_3$. Although $Q_1^2=Q_2^2$ in both cases, the
antisymmetry of the $\epsilon$-tensor does not produce a cancelation of the
two divergences within the reggeon vertex $\Gamma_6$. However, in the 
full hatched diagram of Fig.~4 the $k_3$ integration is symmetric and there
will be a cancelation after integration. 

In higher-orders the two gluons in each $t_i$ channel are replaced by
even signature two-reggeon states with couplings $G_1, G_2$ and $G_3$ to 
the external scattering states and the integration in individual diagrams 
is not necessarily symmetric. However, 
if we keep the same quark loop interaction,
the amplitude with even signature in each channel is obtained 
by summing over all diagrams related by a twist ($P_i \leftrightarrow 
P_{i'}, ~i=1,2,3$). 
Both the color factors and all three of the $k_i$-integrations are then 
symmetric and the anomaly cancels~\cite{arw99}. Sufficient antisymmetry 
to avoid cancelation would appear only if all three reggeon states had 
anomalous color parity ($\neq$ signature). However, such reggeon states do not 
couple to elementary scattering quarks (or gluons), 

In more complicated scattering processes anomaly interactions can appear 
without cancelation. However, all the reggeon states involved carry
anomalous color parity~\cite{arw99}. Such states 
appear when additional particles are produced (or absorbed) at the external
vertices. Since anomaly vertices do not conserve color parity they 
have to appear pairwise. Also, while there is an infra divergence in 
the reggeon interaction in which it appears, the anomaly 
not produce divergences 
in full amplitudes because of compensating reggeon Ward identity 
zeroes~\cite{arw98}.

If some 
external couplings are chosen so that 
particular reggeon Ward identity zeroes are absent,
the anomaly does produce divergent amplitudes. A
reggeon condensate with the quantum numbers of the 
winding-number current can be introduced this way. In \cite{arw98} 
it was argued that in the color superconducting supercritical phase 
this condensate is consistently 
reproduced in all reggeon states by anomaly infra-red 
divergences, while also producing confinement, chiral 
symmetry breaking and a regge pole pomeron. 
Having the full structure of the anomaly under control, 
we hope to implement the program of \cite{arw98} in detail in future papers. 
In effect we aim to to show that in massless QCD
infra-red divergences produce a
transition from perturbative reggeon diagrams 
to diagrams containing hadrons and the 
pomeron. 
If we obtain a unitary (reggeon) S-Matrix, as we anticipate, it will
be very close to perturbation 
theory, with the non-perturbative properties of confinement and chiral 
symmetry breaking a consequence of the anomaly only. 

\newpage

\noindent { \bf References}

\end{document}